\documentstyle[prd,tighten,aps]{revtex}

\begin{document}
\draft

\twocolumn[\hsize\textwidth\columnwidth\hsize\csname
@twocolumnfalse\endcsname

\title{Canonical Quantization of the Gowdy Model}
\author{Guillermo A. Mena Marug\'an} 
\address{Instituto de Matem\'aticas y F\'{\i}sica
Fundamental, C.S.I.C.,\\ Serrano 121, 28006 Madrid, Spain.}
\maketitle

\vspace*{-4.5cm}\begin{flushright}
gr-qc/9704041
\end{flushright}\vspace*{3.3cm}

\begin{abstract}

The family of Gowdy universes with the spatial topology of a 
three-torus is studied both classically and quantum mechanically.
Starting with the Ashtekar formulation of Lorentzian general 
relativity, we introduce a gauge fixing procedure to remove almost 
all of the non-physical degrees of freedom. In this way, we arrive 
at a reduced model that is subject only to one homogeneous 
constraint. The phase space of this model is described by means of a 
canonical set of elementary variables. These are two real, 
homogeneous variables and the Fourier coefficients for four real 
fields that are periodic in the angular coordinate which does not 
correspond to a Killing field of the Gowdy spacetimes. We also obtain
the explicit expressions for the line element and reduced 
Hamiltonian. We then proceed to quantize the system by representing
the elementary variables as linear operators acting on a vector 
space of analytic functionals. The inner product on that space is
selected by imposing Lorentzian reality conditions. We find the
quantum states annihilated by the operator that represents the
homogeneous constraint of the model and construct with them the
Hilbert space of physical states. Finally, we derive the general
form of the quantum observables of the model.

\end{abstract}


\vskip2pc]

\renewcommand{\thesection}{\Roman{section}}
\renewcommand{\thesubsection}{\Alph{subsection}}
\renewcommand{\theequation}{\arabic{section}.\arabic{equation}}
\renewcommand{\thefootnote}{\arabic{footnote}}

\section {Introduction}

The alternative formalism for general relativity put forward by 
Ashtekar [1,2] has renewed the hopes of consistently quantizing 
the gravitational interaction in a non-perturbative way. In contrast
to the situation found in the geometrodynamic formulation, the 
gravitational constraints acquire a simple, polynomic form in terms 
of the Ashtekar canonical variables. Besides, by shifting the 
emphasis from geometrodynamics to connection dynamics, the 
introduction of the Ashtekar variables has allowed the use in gravity 
of mathematical techniques that had been developed in the 
quantization of gauge field theories.

In order to gain insight into the kind of problems that one will 
probably have to face when quantizing full general relativity, a lot 
of attention has been devoted in the last years to the quantization 
of gravitational models with different types of spacetime 
symmetries [3,4]. Most of the systems studied are however 
minisuperspace models [3]. These are clearly inadequate to discuss 
the difficulties that will presumably arise in the quantization 
of full gravity owing to the presence of an infin\-ite number
of degrees of freedom. A possible way to analyze such difficulties 
would be to consider the quantization of midisuperspace models. 
The symmetry of this type of models is not large enough as to 
eliminate all the local degrees of freedom, so that their 
quantization will lead to a true quantum field theory.

In a recent paper [4], Ashtekar and Pierri carried out the 
quantization of the Einstein-Rosen cylindrically symmetric 
spacetimes [5], completing previous works on the subject by 
Kucha\v{r} [6] and Allen [7]. To our knowledge, this is the only 
gravitational midisuperspace model that has been rigorously 
quantized in the literature. 

It would be of interest to have at our disposal other examples 
of midi\-superspace models whose quantization can be achieved.
Natural candidates for such models are provided by spacetimes with 
two commuting spacelike Killing fields [8]. These spacetimes can 
generally be described by two local physical degrees of freedom 
which depend only on one of the spatial coordinates. Besides, since 
the pioneering work by Geroch [9], it is known that the Einstein 
equations of these spacetimes present an infinite number of 
symmetries. It is therefore believed that such systems may in fact 
be classically integrable, because there should exist a conserved 
charge associated with each of the symmetries of the Geroch 
group [10]. Thus, these systems seem to be simple enough as to
expect that their quantization may be feasible.

On the other hand, the existence of the Geroch symmetries is on the 
basis of a series of solution-generating techniques [11] that have 
been developed from different points of view to obtain new solutions 
to the Einstein equations. Thanks to these techniques, it has been 
possible to find a variety of physically interesting classical 
spacetimes with two commuting Killing fields.

Actually, a particular family of spacetimes of this kind is given by
the Einstein-Rosen solutions considered in Ref. [4]. In these 
solutions, the sections of constant time are non-compact, and the 
Killing fields are hypersurface orthogonal. In this paper, we will 
focus our attention on spacetimes which, by contrast, have closed 
spacelike hypersurfaces and whose commuting spacelike Killing fields 
are, in general, not orthogonal. The global structure of the 
spacetimes with these properties has been studied by Gowdy [12], who
has shown that, in this case, the sections of constant time must be
homeomorphic to either $S^1\times S^2$ (a three-handle), a 
three-sphere, or a three-torus (or to a manifold covered by one of 
the above). Among these possible spatial topologies, we will limit 
our discussion exclusively to the case of a three-torus.

A partial symmetry reduction of this Gowdy model can be found in 
Ref. [13]. Preliminary  studies of its quantization, assuming the 
orthogonality of the two Killing fields, have been carried out by 
Berger [14]. In addition, Husain [15,16] has recently proposed gauge
fixing conditions in the Ashtekar formulation for removing all 
the non-dynamical degrees of freedom of the model.
However, he has not performed the gauge fixing to completion.
On the other hand, Husain has not addressed
the quantization of the system in Refs. [15,16]. Our purpose here 
is to complete the gauge fixing procedure and construct a quantum 
framework for the description of the family of Gowdy cosmologies 
with the spatial topology of a three-torus. 

The paper is organized as follows. Sec. II deals with the two 
commuting spacelike Killing field reduction of the Ashtekar formalism
for the case of the Gowdy cosmologies. In that section, we also
present our model and display the expressions of the first-class 
constraints. In Sec. III we introduce a set of gauge fixing 
conditions and show that they are well-posed and consistent. The 
final result of our gauge fixing is that we can remove all the 
first-class constraints, except for a homogeneous one. This 
homogeneous constraint is analogous to the periodicity condition 
discovered by Gowdy [12]. The classical reduced model determined by 
our gauge fixing conditions is studied in Sec. IV. We prove 
that the phase space of the model can be described by using a 
canonical set of elementary variables that are all real. These are 
given by four functions on $S^1$ and two homogeneous 
variables. In addition, we explicitly obtain the metric and the
reduced Hamiltonian that generates the dynamical evolution. Sec. V
is devoted to the quantization of the above reduced model following
the canonical quantization program ellaborated by Ashtekar [2]. 
We first choose a representation space for the quantum theory and 
select an inner product on it by imposing reality conditions [2,17]. 
The homogeneous constraint of the system is then imposed \`a la 
Dirac. The kernel of the quantum constraint provides us with the 
Hilbert space of physical states. In Sec. VI, we determine the form 
of the quantum observables of the reduced model and discuss the 
quantum evolution. Finally, Sec. VII contains the conclusions and 
some further comments.

\section{The Gowdy Model}
\setcounter{equation}{0}

The Gowdy universes are four-dimensional vacuum spacetimes with 
compact spacelike hypersurfaces and two commuting spacelike Killing 
fields [12]. In this paper, we are going to analyze only the case in 
which the spatial topology is that of a three-torus. Besides, we will 
restrict our considerations to non-degenerate Lorentzian metrics.

Let us first introduce the Ashtekar formalism for Lorentzian general
relativity, particularizing then to the Gowdy model. The Ashtekar 
gravitational variables can be taken as a densitized triad, 
$\tilde{E}^a_i$, and a SO(3) connection, $A_a^i$, both defined on a
three-manifold $\Sigma$ [2]. Lower case Latin letters from the 
beginning and the middle of the alphabet denote spatial and SO(3)
indices, respectively. The SO(3) indices run from 1 to 3, and are
raised and lowered with the metric $\eta_{ij}={\rm diag}(1,1,1)$.
For Lorentzian gravity, the Poisson bracket structure is given by
\begin{equation} \{A_a^i(x),\tilde{E}^b_j(y)\}=i\delta_a^b 
\delta^i_j\delta^{(3)}(x-y).\end{equation}
In this formula, $x$ and $y$ are two generic points of $\Sigma$, 
$\delta^b_a$ is the Kronecker delta, and $\delta^{(3)}$ is the 
delta function on $\Sigma$.

Provided that the metric is non-degenerate, the Ashte\-kar variables 
can be expressed in terms of the triad, $e^a_i$, and the extinsic 
curvature [18], $k_{ab}$,
\begin{equation} \tilde{E}^a_i=e^a_i h(e),\;\;\;\;\;
A_a^i=\Gamma_a^i(e) -i k_{ab} e^{bi},\end{equation}
where $h=({\rm det}\,h_{ab})^{1/2}$, $h^{ab}=e^a_ie^{bi}$ is the
inverse three-metric, and $\Gamma_a^i$ is the SO(3) connection 
compatible with the triad [19],
\begin{equation} \Gamma_a^i=-\frac{1}{2} \epsilon^{ijk} 
E_{_{_{\!\!\!\!\!\!\sim}}\;jb} (\partial_a\tilde{E}^b_k+
\Gamma^b_{\;ca}\tilde{E}^c_k).\end{equation}
Here, $\epsilon^{ijk}$ is the antisymmetric symbol, 
$E_{_{_{\!\!\!\!\!\!\sim}}\;ia}$ the inverse of the densitized triad,
and $\Gamma^a_{\;bc}$ the Christoffel symbols [18].

In the Ashtekar formalism, the first-class constraints of vacuum
general relativity are [2]
\begin{eqnarray}{\cal G}_i & \equiv & {\cal D}_a \tilde{E}^a_i=
\partial_a \tilde{E}^a_i
+ \epsilon_{ij}^{\;\;\;\;k} A_a^j \tilde{E}^a_k=0,\\
{\cal C}_a & \equiv & F_{ab}^i \tilde{E}^b_i=0,\\
 {\cal H} & \equiv & \epsilon_i^{\;\;jk}F_{ab}^i \tilde{E}^a_j
\tilde{E}^b_k=0,\end{eqnarray}
where $F_{ab}^i$ is the curvature of the SO(3) connection,
\begin{equation} F_{ab}^i=\partial_aA_b^i-\partial_bA_a^i+
\epsilon^i_{\;\;jk}A_a^jA_b^k.\end{equation}

For the Gowdy universes with the topology of a three-torus, we can 
always choose spatial coordinates $\omega$, $\nu$ and $\theta$ such 
that $(\partial_\omega)^a$ and $(\partial_\nu)^a$ are the two 
commuting Killing fields. For later convenience, we will normalize 
the periods of these coordinates so that 
$2\pi\omega,\;2\pi\nu,\;\theta\in S^1$. All variables of the model 
must then depend only on $\theta$ and the time coordinate, $t$. 
Furthermore, they have to be periodic in $\theta\in S^1$.

On the other hand, following Husain and Smolin [13], we can set equal
to zero the densitized triad components
\begin{equation} \tilde{E}^{\theta}_1=\tilde{E}^{\theta}_2=
\tilde{E}^{\omega}_3=\tilde{E}^{\nu}_3=0. \end{equation}
The constraints ${\cal G}_1$, ${\cal G}_2$, ${\cal C}_{\omega}$ and 
${\cal C}_{\nu}$ are then solved by
\begin{equation} A_{\theta}^1=A_{\theta}^2=A_{\omega}^3=A_{\nu}^3=0.
\end{equation}
After this symmetry reduction, and renaming $A= A_{\theta}^3$, 
$E= \tilde{E}^{\theta}_3$, the remaining first-class constraints of
the system can be written [15]
\begin{eqnarray} G & \equiv & \partial_{\theta}E+J=0,\\
 C & \equiv & \tilde{E}^{\alpha}_L\partial_{\theta}
A_{\alpha}^L+AJ=0,\\
H & \equiv & 2E \tilde{E}^{\alpha}_ L \epsilon^{L}_{\;\;M}
\partial_{\theta}A_{\alpha}^{M}+2AEK\nonumber\\
&&-K_{\alpha}^{\;\beta}
K_{\beta}^{\;\alpha}+K^2=0,\end{eqnarray}
where $\alpha,\beta\!=\!\omega$ or $\nu$, $\;L,M\!=\!1$ or 2, 
$\epsilon^{LM}$ is the antisymmetric symbol in two dimensions, and 
we have employed the notation
\begin{eqnarray} & K_{\alpha}^{\;\beta} =  A_{\alpha}^I
\tilde{E}^{\beta}_I,
\;\;\;\;\;\;\;\;\;\;\;\;& K = K_{\alpha}^{\;\alpha},\\
& J_{\alpha}^{\;\beta} = \epsilon_{L}^{\;\;M}A_{\alpha}^{L}
\tilde{E}^{\beta}_M\;,\;\;\;\;\;& J = 
J_{\alpha}^{\;\alpha}.\end{eqnarray}

Thus, our Gowdy model can be described by the ten fields 
$(A,E,A_{\alpha}^L,\tilde{E}^{\alpha}_L)$, which are periodic 
functions of $\theta$. The Lorentzian symplectic structure is 
determined by the Poisson brackets
\begin{eqnarray} &&\{A(\theta),E(\theta^{\prime})\}=i\delta(\theta
-\theta^{\prime}),\\ 
&&\{A_{\alpha}^L(\theta),
\tilde{E}^{\beta}_M(\theta^{\prime})\}=i\delta_{\alpha}^{\beta}
\delta_M^L \delta(\theta-\theta^{\prime}),\end{eqnarray} 
$\delta(\theta)$ being the delta function on $S^1$.
These fields are subject to the constraints (2.10-12), which will 
be referred from now on as the Gauss, diffeomorphism, 
and scalar constraint, respectively. Their physical interpretation 
and Poisson algebra has been discussed by Husain [15,16]. 

It is worth noting that the variables $K_{\alpha}^{\;\beta}$
and $J_{\alpha}^{\;\beta}$ are not functionally independent, 
for one can check that 
${\rm det}\, K_{\alpha}^{\;\beta}={\rm det}\, J_{\alpha}^{\;\beta}$.
Therefore, one cannot replace the elementary variables
$(A_{\alpha}^L,\tilde{E}^{\alpha}_L)$ with the eight Gauss-invariant
quantities $(K_{\alpha}^{\;\beta},J_{\alpha}^{\;\beta})$.

It will prove most convenient to introduce instead a change of phase
space variables from $(A_{\alpha}^L,\tilde{E}^{\alpha}_L)$ to
$K_{\omega}^{\;\omega}$, $K_{\omega}^{\;\nu}$, $K$, $J$, and
\begin{eqnarray}   &&
x=\frac{q^{\omega\omega}}{q^{\nu\nu}},\hspace{1.6cm}
v=\frac{q^{\omega\nu}}{q^{\nu\nu}},\\
&&w=\frac{1}{2}\ln{q^{\nu\nu}},\hspace{.9cm}
\phi=\arctan{\left(\frac{\tilde{E}^{\nu}_1}
{\tilde{E}^{\nu}_2}\right)},\end{eqnarray}
where
\begin{equation} q^{\alpha\beta}=
\tilde{E}^{\alpha}_L\tilde{E}^{\beta L}.\end{equation}
From Eqs. (2.2) and (2.8), we get that 
$q^{\alpha\beta}=h^{\alpha\beta} h^2$, so that, for positive 
definite three-metrics, $q^{\alpha\beta}$ must also be positive 
definite. Therefore, $x$, $v$ and $w$ are well-defined by
Eqs. (2.17,18), and we must have $x>0$, $v,w\in I\!\!\!\,R$, and 
$x>v^2$, this last inequality coming from the fact that
\begin{equation} {\rm det}\, q^{\alpha\beta}= e^{4w} (x-v^2)>0.
\end{equation}
As to the variables $K_{\omega}^{\;\omega}$, $K_{\omega}^{\;\nu}$, 
$K$, $J$, and $\phi$, we will admit for the moment that they are 
complex.

Let us show that the above change of variables can always be inverted
in the sector of positive definite three-metrics. Using Eq. (2.19), 
relations (2.18) can be equivalently written in the form
\begin{equation} \tilde{E}^{\nu}_1=e^w \sin{\phi},\;\;\;\;\;
\tilde{E}^{\nu}_2=e^w \cos{\phi}.\end{equation}
The definitions of $x$ and $v$ lead in turn to
\begin{eqnarray}&&\tilde{E}^{\omega}_1=v\tilde{E}^{\nu}_1+
\sqrt{x-v^2}\tilde{E}^{\nu}_2,\\
&&\tilde{E}^{\omega}_2=v\tilde{E}^{\nu}_2-\sqrt{x-v^2}
\tilde{E}^{\nu}_1.\end{eqnarray}
So, given $v,w\in I\!\!\!\,R$, $x>v^2$, and $\phi$, one can always 
recover $\tilde{E}^{\alpha}_L$. Besides, Eqs. (2.13,14) can be seen 
to imply the identities
\begin{eqnarray} & K_{\nu}^{\;\nu}& = K-K_{\omega}^{\;\omega},\\
& K_{\nu}^{\;\omega}& = (K-2K_{\omega}^{\;\omega})v+
K_{\omega}^{\;\nu}x+J \sqrt{x-v^2},\end{eqnarray}
that enable us to find the missing components of 
$K_{\alpha}^{\;\beta}$ from our new variables. Once 
$K_{\alpha}^{\;\beta}$ and $\tilde{E}^{\alpha}_L$ are known, we can 
finally obtain $A_{\alpha}^L$ through
\begin{equation} A_{\alpha}^L=K_{\alpha}^{\;\beta}
E_{_{_{\!\!\!\!\!\!\sim}}\;\beta}^L,\end{equation}
$E_{_{_{\!\!\!\!\!\!\sim}}\;\alpha}^L$ being the inverse of 
$\tilde{E}^{\alpha}_L$, which can always be computed because 
$q^{\alpha\beta}$ is positive definite.

As far as we restrict our attention to the sector of Lorentzian 
non-degenerate metrics, the variables introduced above, together 
with $A$ and $E$, can then be regarded as a set of elementary 
variables for our model. Moreover, it is easy to check from 
Eqs. (2.15,16) that they form a closed Poisson algebra. The only 
non-vanishing brackets are
\begin{eqnarray} \{A(\theta),E(\theta^{\prime})\}=&\{J(\theta),\phi 
(\theta^{\prime})\}&=\{K(\theta),w(\theta^{\prime})\}\nonumber\\
&=i\delta(\theta-\theta^{\prime}),&\end{eqnarray}
\begin{eqnarray}&&\{K_{\omega}^{\;\omega}(\theta),
K_{\omega}^{\;\nu}(\theta^{\prime})\}=-i K_{\omega}^{\;\nu}(\theta)
\delta(\theta-\theta^{\prime}),\\
&&\{K_{\omega}^{\;\omega}(\theta),x(\theta^{\prime})\}=
2i x(\theta) \delta(\theta-\theta^{\prime}),\\
&&\{K_{\omega}^{\;\omega}(\theta),v(\theta^{\prime})\}=
i v(\theta) \delta(\theta-\theta^{\prime}),\\
&& \{K_{\omega}^{\;\nu}(\theta),x(\theta^{\prime})\}=
2i v(\theta) \delta(\theta-\theta^{\prime}),\\
&&\{K_{\omega}^{\;\nu}(\theta), v(\theta^{\prime})\}=i 
\delta(\theta-\theta^{\prime}).\end{eqnarray}

\section{Gauge Fixing}
\setcounter{equation}{0}

We will now eliminate non-physical degrees of freedom from our set
of phase space variables by introducing suitable gauge fixing
conditions. These conditions, together with the 
constraints (2.10-12), will provide us with a set of 
second-class constraints that will allow to reduce the model. The 
gauge fixing conditions that we are going to impose are
\begin{eqnarray} \chi_H & \equiv & E-e^t=0,\\
 \chi_G & \equiv & \phi=0,\\
 \chi_C & \equiv & K-\frac{K_0}{\sqrt{2\pi}}=0,\end{eqnarray}
where
\begin{equation} K_0=\oint \frac{K}{\sqrt{2\pi}}.\end{equation}
Here, the symbol $\oint$ denotes integration over $\theta\in S^1$.

In Eq. (3.1), the time coordinate $t$ is assumed to be real. This
condition will be seen to fix the gauge freedom associated with the
scalar constraint (2.12). Our gauge fixing is in fact equivalent to
Gowdy's choice of time [12] (and, therefore, to that made  
in Ref. [16]), which can be expressed as $E=\tau$, $\tau$
being a strictly positive time coordinate. None the less, we
notice that, while in Gowdy's time all classical solutions present 
a cosmological singularity at $\tau=0$ [12], this singularity is 
driven to minus infinity with our choice of gauge. In this way, we
allow a domain of definition for $t$ that is the whole real axis.

We will also prove that the requirement (3.2) fixes completely the 
Gauss gauge of our model. From Eq. (2.18), this requirement implies 
$\tilde{E}^{\nu}_1=0$. Finally, we will show that condition (3.3) 
(which was employed in Ref. [16]) removes almost entirely the 
diffeomorphism gauge freedom. Some comments are in order concerning 
the appearance of $K_0$ in this gauge fixing condition. The quantity
$K_0$ is known to be a classical Dirac observable of the system 
[20],in the sense that its Poisson brackets with all the first-class
constraints (2.10-12) vanish weakly. As a consequence, $K_0$ is a 
constant of motion whose value depends only on the particular 
solution that is being considered. This value is invariant under any
gauge transformation. On the other hand, since $K$ is a periodic 
function of $\theta$, it must admit a Fourier series of the 
form\footnote{We assume that all the classical elementary variables,
and in particular $K$, are smooth functions of $\theta$. In fact, 
it suffices that $K\in C^1(S^1)$ for its Fourier series to converge 
to $K$ at all points $\theta\in S^1$.}
\begin{equation} K=\sum_{n=-\infty}^{\infty} K_n(t) \,
\frac{e^{in\theta}}{\sqrt{2\pi}},\;\;\;\;\;\;\;\;
K_n(t)=\oint K \,\frac{e^{-in\theta}}{\sqrt{2\pi}}.\end{equation}
Condition (3.3) amounts thus to absorb all the Fourier coefficients
$K_n$ with $n\neq 0$ by means of a diffeomorphism.

Let us now see that our gauge fixing conditions are well-posed. A
straightforward calculation shows that
\begin{eqnarray} && \{\chi_H,\oint n_{_{_{\!\!\!\!\!\sim}}} H\} 
= -2i n_{_{_{\!\!\!\!\!\sim}}\;}EK,\hspace{1.cm}\\
&& \{\chi_G,\oint \lambda  G\} = -i\lambda,\\
&& \{\chi_C,\oint n  C\}=i\partial_{\theta}(nK),\end{eqnarray}
where $\lambda$ and $n$ are functions on $S^1$ and 
$n_{_{_{\!\!\!\!\!\sim}}\;}$ is a density of weight $-1$.
If $n_{_{_{\!\!\!\!\!\sim}}\;}$, $\lambda$ and $\partial_{\theta}n$ 
are different from zero, conditions (3.1) and (3.3) guarantee that 
these Poisson brackets never vanish for $K_0\neq 0$. Therefore, 
provided that $K_0$ does not vanish, our gauge fixing conditions are 
second-class with the constraints, and hence acceptable.

The problems found at $K_0=0$ can be obviated in the following sense.
Using Eqs. (2.2,3), (2.8,9), and (2.13), it is possible to show that 
the variable $K$ can be equivalently expressed in our model as
\begin{equation} K=-i\, h\, k_{\alpha\beta} h^{\alpha\beta}.
\end{equation}
Here, $h$ is again the square root of the determinant of the 
three-metric, and $k_{\alpha\beta}$ and $h^{\alpha\beta}$ denote, 
respectively, the ($\alpha\beta$)-components of the extrinsic 
curvature and the inverse three-metric (with 
$\alpha, \beta=\omega$ or $\nu$). Then, $K$ must be purely imaginary 
if the three-metric is positive definite. Suppose now that $K_0=0$. 
Since $K$ is imaginary and periodic, it follows that it must
vanish at least at one point $\theta_0\in S^1$ on each section of 
constant time. But one can then easily check that all Poisson 
brackets of $\chi_H$ with the first-class constraints vanish at 
$\theta_0$, modulo such constraints and our gauge fixing condition. 
So, our gauge fixing is not admissible if $K_0=0$. The same 
conclusion is reached if one adopts Gowdy's choice of time, 
$E=\tau$. As a consequence, the classical solutions with $K_0=0$, 
that are not compatible with our gauge fixing, turn out not to be
included in the family of cosmologies with the topology of a 
three-torus studied by Gowdy [12]. Since we are only interested in
analyzing this family of solutions, we can disregard the case 
$K_0=0$. Furthermore, we will see in Sec. IV that the geometry of 
these solutions can be considered invariant under a change of sign 
in $K_0$. Making use of this symmetry, we can set 
$iK_0\in I\!\!\!\,R^+$ without loss of generality. In this way, the
point $K_0=0$ will be driven to the boundary of our reduced phase
space. Under quantization, the possible inclusion of that point will
be physically irrelevant inasmuch as it will correspond to a set of
measure zero in the phase space of the system.

From now on, we will thus take $K_0\neq 0$. Equations (3.6-8) ensure 
then that our conditions (3.1-3) are suitable to fix the scalar, 
Gauss and diffeomorphism gauge degrees of freedom. On the other hand, 
employing our gauge fixing conditions, we can solve the scalar and 
Gauss constraints to obtain the expressions for $A$ and $J$ as 
functions of the variables 
$K_{\omega}^{\;\omega}$, $K_{\omega}^{\;\nu}$, $x$, $v$, and $K_0$:
\begin{eqnarray}A\!&=&\!\frac{\sqrt{2\pi}}{2K_0\sqrt{x-v^2}}
(K_{\omega}^{\;\nu}\partial_{\theta}x-2K_{\omega}^{\;\omega}
\partial_{\theta}v)\!+\!e^{-t}(K_{\omega}^{\;\nu}v-
K_{\omega}^{\;\omega}) \nonumber\\
& &+\frac{\sqrt{2\pi}e^{-t}}{K_0}[(K_{\omega}^{\;\omega}
-K_{\;\omega}^{\;\nu}v)^2
+(K_{\omega}^{\;\nu})^2(x-v^2)],\end{eqnarray}
\begin{equation} J=0.\end{equation}
This and Eqs. (3.1,2) remove the two canonically conjugate
pairs $(A,E)$ and $(J,\phi)$ as dynamical degrees of freedom.

In addition, the diffeomorphism constraint (2.11) can now be 
rewritten
\begin{equation} \Pi^{\prime}-\frac{K_0}{\sqrt{2\pi}} 
\partial_{\theta}w=0,\end{equation}
where
\begin{equation}\Pi^{\prime}=\frac{1}{2(x-v^2)}[K_{\omega}^{\;\nu}
(v\partial_{\theta}x-2x\partial_{\theta}v)-
K_{\omega}^{\;\omega}\partial_{\theta}(x-v^2)].\end{equation}
Since our fields have a periodic dependence on the angular 
coordinate $\theta$, $\Pi$ and $w$ can be expanded as Fourier series
similar to that displayed for $K$ in Eq. (3.5). Formula (3.12) fixes
then all the Fourier coefficients $w_n$ with $n\neq 0$ in terms of 
$K_0$ and the Fourier coefficients of $\Pi^{\prime}$,
\begin{equation} w_n=\frac{\sqrt{2\pi}\Pi^{\prime}_n}{inK_0},
\;\;\;\;\;\;\;n\neq 0.\end{equation}
The coefficient $w_0$ is however left undetermined. Besides, 
integration over $S^1$ of Eq. (3.12) leads to the global constraint
\begin{equation} \Pi^{\prime}_0=\oint 
\frac{\Pi^{\prime}}{\sqrt{2\pi}}=0.\end{equation}

We recall at this point that our gauge fixing condition (3.3) 
amounts to set all the Fourier coefficients $K_n$ of $K$, except 
$K_0$, equal to zero. On the other hand, Eqs. (2.27-32) imply that 
$K_n$ and $w_n$ commute under Poisson brackets with the rest of our
phase space variables, whereas
\begin{equation} \{K_n,w_{m}\}=i\delta_{-n}^{m}.\end{equation}
We conclude in this way that our gauge fixing condition, together 
with the diffeomorphism constraint, allow us to eliminate the 
canonically conjugate pairs $(K_n,w_{-n})$ with $n\neq 0$ as 
physical degrees of freedom, while the homogeneous components of 
$K$ and $w$ (i.e. the Fourier coefficients $K_0$ and $w_0$) 
remain as dynamical variables. In this reduction process, the 
diffeomorphism gauge freedom is not totally removed, because we 
are still left with the homogenous part of the diffeomorphism 
constraint, $\Pi^{\prime}_0=0$.

In order to prove that our gauge fixing procedure is consistent, we 
still have to show that the conditions (3.1-3) are compatible with 
the dynamical evolution of the model. This evolution is generated by 
the total Hamiltonian constraint [2]
\begin{equation} H^T=\oint
\left[-\frac{N_{_{_{\!\!\!\!\!\!\sim}}\;}}{2}
H-iN^{\theta}(C-AG)-i\Lambda G\right],\end{equation}
where $H$, $C$, and $G$ are the first-class constraints (2.10-12),
$N_{_{_{\!\!\!\!\!\!\sim}}\;}$ is the densitized lapse function, 
$N^{\theta}$ is the only non-vanishing component of the shift 
vector,\footnote{The $\omega$ and $\nu$ components of the shift 
vector can be made equal to zero after the symmetry 
reduction (2.8,9).} and $\Lambda$ is a Lagrange multiplier. Besides, 
$N_{_{_{\!\!\!\!\!\!\sim}}\;}$ and $N^{\theta}$ are real if the
metric is Lorentzian, and $N_{_{_{\!\!\!\!\!\!\sim}}\;}$ must be 
different from zero. What we have to check then is that there exists 
a choice of densitized lapse, shift, and $\Lambda$ such that the 
total time derivative of each of our gauge conditions vanishes. 
This total time derivative (that will be denoted by a dot) is given
by the sum of the Poisson bracket with $H^T$, $\{\;.\;,H^T\}$, and 
the partial derivative with respect to the explicit dependence on the
time coordinate, $\partial_t$. After a careful calculation, we get 
that, modulo constraints and gauge fixing conditions,
\begin{eqnarray} \dot{\chi}_H&=&
e^t\left(iN_{_{_{\!\!\!\!\!\!\sim}}\;}
\frac{K_0}{\sqrt{2\pi}}-1\right),\\
\dot{\chi}_G&=&-\Lambda-iN_{_{_{\!\!\!\!\!\!\sim}}\;}
\left(\sqrt{x-v^2}K_{\omega}^{\;\nu}-\frac{\sqrt{2\pi}e^t}
{K_0}\Pi^{\prime}\right)\nonumber\\
&&+ie^t\partial_{\theta}N_{_{_{\!\!\!\!\!\!\sim}}\;},\\
\dot{\chi}_C&=&\frac{K_0}{\sqrt{2\pi}}\partial_{\theta}
N^{\theta}.\end{eqnarray}

The requirement that $\dot{\chi}_H$ vanishes implies
\begin{equation} N_{_{_{\!\!\!\!\!\!\sim}}\;}=\frac{\sqrt{2\pi}}
{iK_0}.\end{equation}
This and $\dot{\chi}_G=0$ determine a unique $\Lambda$ through 
Eq. (3.19). Finally, by demanding that $\dot{\chi}_C=0$, we conclude 
that $N^{\theta}$ can be any function of $t$, 
$N^{\theta}=N^{\theta}(t)$. We thus see again that the diffeomorphism
gauge freedom has not been entirely removed, since the shift 
function $N^{\theta}$ is not completely fixed. Any diffeomorphism 
with infinitesimal parameter $N^{\theta}(t)$ is still allowed. Note 
that such diffeomorphisms are precisely those generated by the only
remaining constraint $\Pi^{\prime}_0=0$. 

For Lorentzian 
metrics, the shift $N^{\theta}(t)$ must be real. On the other hand, 
we have commented above that $K$, and hence $K_0$ [from Eq. (3.4)],
is purely imaginary in the Lorentzian case. Therefore, the densitized
lapse (3.21) is actually real. Moreover, it does not vanish for any 
finite value of $K_0$. This concludes the proof of consistency of our
gauge fixing conditions with the Lorentzian evolution.

The final result of our gauge fixing procedure is the elimination 
of the non-dynamical fields ($A,E,J,\phi$) and Fourier coefficients 
($K_n,w_n$) with $n\neq 0$. The phase space of the reduced
model can be described by the four (periodic) fields 
$K_{\omega}^{\;\omega}$, $K_{\omega}^{\;\nu}$, $x$, and $v$, and the
two homogeneous variables $K_0$ and $w_0$. Since all these variables
commuted with the non-physical degrees of freedom that have been 
suppressed, the reduction of the system does not alter their Poisson
brackets (i.e., their Poisson and Dirac brackets coincide). 
These brackets are given by Eqs. (2.28-32) and
\begin{equation} \{K_0,w_0\}=i,\end{equation}
which follows from Eq. (3.16). Finally, the reduced model is still 
subject to one homogeneous constraint, namely, $\Pi^{\prime}_0=0$.

\section{The Reduced Classical Model}
\setcounter{equation}{0}

In this section, we will analyze the dynamical evolution of our
reduced model, discuss the reality conditions on the phase space
variables, and obtain the expression of the classical metric.

\begin{center}
\subsection{Classical evolution}
\end{center}

We have seen that the shift function $N^{\theta}$ can be any function
of time, $N^{\theta}(t)$. For practical purposes, none the less, this
shift function can always be absorbed by replacing the angular
coordinate $\theta$ with
\begin{equation} \theta^{\prime}=\theta+\int_{t_0}^t dt^{\prime} 
N^{\theta}(t^{\prime}),\end{equation}
where $t_0$ is any given time. From the periodicity of $\theta$, it 
follows that $\theta^{\prime}$ is defined as well on $S^1$. Notice 
also that, for fixed $t$, 
$\partial_{\theta^{\prime}}=\partial_{\theta}$.

Since the vector fields $(\partial_{\theta^{\prime}})^{\mu}$ and 
$(\partial_t)^{\mu}$ ($\mu=t,a$) are orthogonal, the dynamical 
evolution (with $\theta^{\prime}$ kept constant) is generated in our 
model by the reduced Hamiltonian density, $H_R$, which is provided 
by the negative of the momentum canonically conjugate to the 
variable chosen as time. Recalling condition (3.1), which implies
that $t=\ln E$, and taking into account the Lorentzian Poisson
brackets (2.27-32), we conclude that the momentum canonically 
conjugate to our time variable is given by $iAE$. In this way, we 
arrive at $H_R=-iAe^t$, with $A$ given by Eq. (3.10). 
Hence, for constant $\theta^{\prime}$, the 
time derivative of any of our reduced phase space variables, f, is 
\begin{equation} \dot{f}=\partial_t f+\{f,\oint H_R\}.\end{equation}
In the following, we will assume that the shift $N^{\theta}(t)$ has
already been absorbed in $\theta^{\prime}$ and suppress the prime 
from this angular coordinate.

\begin{center}
\subsection{Real phase space variables}
\end{center}

Our variables 
$(x,v,K_{\omega}^{\;\omega},K_{\omega}^{\;\nu},K_0,w_0)$ 
present the problem of possessing domains of definition that are 
rather complicated in the sector of non-degenerate Lorentzian 
metrics. On the one hand, we know that $x$ must be greater than 
$v^2$, $v$ being real. On the other hand, particularizing to our 
gauge-fixed model the definition of $K_{\alpha}^{\;\beta}$ in 
Eq. (2.13) and expressions (2.2,3) for the Ashtekar connection, it 
is possible to show that
\begin{eqnarray} K_{\omega}^{\;\omega}&=&
\frac{e^t}{4(x-v^2)^{\frac{3}{2}}}
(2x\partial_{\theta}v-v\partial_{\theta} x)\nonumber\\ 
&&-i e^{\,\frac{t}{2}\,} 
e^w(x-v^2)^{\frac{1}{4}}k^{\omega}_{\omega},\\
K_{\omega}^{\;\nu}&=&-\frac{e^t}{4(x-v^2)^{\frac{3}{2}}}
\partial_{\theta}(x-v^2)\nonumber\\
&& -i e^{\,\frac{t}{2}\,} e^w
(x-v^2)^{\frac{1}{4}}k_{\omega}^{\nu}.\end{eqnarray}
In these formulas, $k_a^b$ is the extrinsic curvature, and 
$w=\sum_n w_n e^{in\theta}/\sqrt{2\pi}$, with $w_n$ determined
by Eq. (3.14) for all $n\neq 0$. So, 
$K_{\omega}^{\;\omega}$ and $K_{\omega}^{\;\nu}$ are not only 
complex, but their real parts are besides functionally dependent on 
$x$ and $v$ in an explicitly time dependent way. We finally recall 
that, for positive definite three-metrics, the functions $w$ and $K$
have to be real and purely imaginary, respectively. In 
addition, we have assumed that $K_0\neq 0$. Therefore, we must have
$w_0\in I\!\!\!\,R$ and $iK_0\in I\!\!\!\,R^+\cup I\!\!\!\,R^-$.

To overcome the above problems, we will now perform a change to
a different set of elementary variables whose elements are all real.
As a plus, these new variables will form a remarkably simple algebra 
under Poisson brackets.

Our first step will consist in replacing $x$ with a new metric 
variable, $y$, whose domain of definition is the entire real axis,
\begin{equation} y=\ln{(x-v^2)}.\end{equation}
The condition $x>v^2$ ensures that $y$ is real and well-defined.
Using then Eqs. (2.28-32), it is not difficult to show that the 
variables
\begin{equation} P_y=\frac{i}{2}(K_{\omega}^{\;\omega}-
K_{\omega}^{\;\nu}v),
\;\;\;\;\;\;\;\;P_v=iK_{\omega}^{\;\nu}\end{equation}
are the momenta canonically conjugate to $y$ and $v$, 
\begin{equation} \{y(\theta),P_y(\theta^{\prime})\}=\{v(\theta),
P_v(\theta^{\prime})\}=\delta(\theta-\theta^{\prime}).\end{equation}
The inverse of relations (4.5,6) is
\begin{eqnarray} x=e^y+v^2,&&\\
K_{\omega}^{\;\omega}=-i(2P_y+vP_v),\;\;\;\;\;\;&&
\;K_{\omega}^{\;\nu}=-iP_v.\end{eqnarray}

On the other hand, it is obvious that
\begin{equation} k_0=iK_0\end{equation}
is real and, given Eq. (3.22), canonically conjugate to $w_0$,
\begin{equation} \{w_0,k_0\}=1.\end{equation} 
Besides, $w_0$ and $k_0$ commute under Poisson brackets with
$y$, $v$, $P_y$, and $P_v$. We have hence attained a canonical set
of elementary variables for our reduced model.

However, the variables $P_y$ and $P_v$ are still complex.
Actually, it follows from Eqs. (4.3,4) and (4.6) that the real parts
of $P_y$ and $P_v$ run over the whole real axis, whereas their 
imaginary parts, ${\cal I}(P_y)$ and ${\cal I}(P_v)$, are restricted 
to be
\begin{equation} {\cal I}(P_y)=\frac{e^t}{4}e^{\,-\frac{y}{2}\,}
\partial_{\theta}v,\;\;\;\;\;{\cal I}(P_v)=-\frac{e^t}{4}
e^{\,-\frac{y}{2}\,} \partial_{\theta}y.\end{equation}
Nevertheless, it is now easy to arrive at a set of real 
elementary variables. This can be achieved by means
of the canonical tranformation generated by the functional
\begin{equation} {\cal F}=w_0k_0^{\prime}+\oint \left[
(y-2t)p_u+vp_v\right] +i F,
\end{equation}
where
\begin{equation}F=-\;\frac{e^t}{2}\oint e^{\,-\frac{y}{2}\,}
\partial_{\theta}v\end{equation}
and $p_u(\theta)$, $p_v(\theta)$, and $k_0^{\prime}$ are the new 
momenta. It is straightforward to check that $k_0=k_0^{\prime}$ and 
that our configuration variables $v(\theta)$ and $w_0$ 
are not affected by the transformation. The field $y(\theta)$ is 
however replaced with a new configuration variable, $u(\theta)$,
whose domain of definition is given again by the real axis:
\begin{equation} u=y-2t.\end{equation}
One can see that this change of variable partly simplifies the 
explicit time dependence of the metric.
More importantly, since $F$ satisfies
\begin{equation} \frac{\delta F}{\delta y}={\cal I}(P_y),\;\;\;\;\;\;
\;\;\frac{\delta F}{\delta v}={\cal I}(P_v),\end{equation}
the new momenta $p_u$ and $p_v$ turn out to coincide with the real
parts of $P_y$ and $P_v$, respectively,
\begin{equation} P_y=p_u+i{\cal I}(P_y),\;\;\;\;\;\;\;\;
P_v=p_v+i{\cal I}(P_v).\end{equation}

Therefore, $(u,p_u,v,p_v)$ (which are fields on $S^1$) and 
$(w_0, k_0)$ provide a canonical set of real elementary variables for
our reduced model. All of these variables run over the whole real 
axis, except $k_0$, which has to be non-vanishing. Besides, the
model possesses one homogeneous constraint, that is given by 
Eq. (3.15). After some calculus, we can express it as
\begin{eqnarray} &&\Pi_0=-i\Pi^{\prime}_0=\oint 
\frac{\Pi}{\sqrt{2\pi}}=0,\\
&&\Pi=-i\Pi^{\prime}=
\partial_{\theta}u\,p_u+\partial_{\theta}v\,p_v.\end{eqnarray}
This constraint reproduces the periodicity condition found by Gowdy
[12]. In fact, it was obtained from Eq. (3.12) by assuming that the
function $w$ is periodic. As remarked by Gowdy, it can also be
interpreted as the condition that the ``total field momentum be 
zero'' [12].

Let us finally notice that, since the generating functional 
${\cal F}$ of the above canonical transformation is explicitly time 
dependent, the dynamical evolution of the variables 
$(u,p_u,v,p_v,w_0,k_0)$ is not generated by the Hamiltonian $H_R$ 
anymore. The new reduced Hamiltonian, $H_r$, can be obtained from 
the standard formula
\begin{equation} \oint H_r
=\oint H_R+\partial_t {\cal F}.\end{equation}
Using that $\partial_t{\cal F}=-2\oint p_u+iF$ and $H_R=-iAe^t$, 
with $A$ given by Eq. (3.10), we get
\begin{eqnarray} H_r & = &
-\frac{\sqrt{2\pi}}{k_0}\left[4p_u^2+ e^{2t}e^u p_v^2
+\frac{e^{2t}}{16}\,(\partial_{\theta}u)^2
+\frac{e^{-u}}{4}\,(\partial_{\theta}v)^2
\right]\!.\nonumber\\
& & \end{eqnarray}
It is straightforward to see that this Hamiltonian 
is bounded from above (below) for positive
(negative) values of $k_0$. 

\begin{center}
\subsection{The metric}
\end{center}

We will now obtain the expression of the classical metric that 
results from our gauge fixing conditions for the Gowdy model. Using 
Eqs. (2.2), (2.8), (2.17-19), and the definition of $u$,
we get that the only non-vanishing components of the three-metric 
$h_{ab}$ are
\begin{eqnarray} &&h_{\theta\theta}= e^{2w} 
e^{\,\frac{u}{2}\,},\;\;\;\;\;\;\;\;\;\;\;\;\;\;\;\;\;\;\;
h_{\alpha\beta}=e^{\,-\frac{u}{2}\,} g_{\alpha\beta},\\
&&g_{\omega\omega}=1,\;\;\;\;\;\;g_{\omega\nu}=-v,
\;\;\;\;\;g_{\nu\nu}=e^{2t}e^u+v^2.\end{eqnarray}
The lapse function, 
$N=N_{_{_{\!\!\!\!\!\!\sim}}\;}({\rm det}\,h_{ab})^{1/2}$, can then
be found from Eq. (3.21),
\begin{equation} N=\frac{\sqrt{2\pi}}{k_0} e^t e^w
e^{\,\frac{u}{4}\,}.\end{equation}
Introducing now the change of time coordinate
\begin{equation} T=\frac{\sqrt{2\pi}}{|k_0|} e^t,\;\;\;\;\;\;\;\;
T\in I\!\!\!\,R^+,\end{equation}
and remembering that the shift function $N^{\theta}$ has been 
absorbed in the angular coordinate $\theta$, we arrive at a line 
element of the form
\begin{eqnarray} ds^2&=& e^{2w} 
e^{\,\frac{u}{2}\,}(-dT^2+d\theta^2)\nonumber\\
&+&e^{\,-\frac{u}{2}\,}
\left(g_{\omega\omega}d\omega^2+2g_{\omega\nu} d\omega d\nu
+g_{\nu\nu}d\nu^2\right)\!,\!\end{eqnarray}
where, in terms of the positive time coordinate $T$, $g_{\nu\nu}$
reads
\begin{equation} g_{\nu\nu}=\frac{k_0^2}{2\pi}\,T^2e^u+v^2.
\end{equation}

On the other hand, Eq. (3.14) determines $w$ to be
\begin{equation} w= \frac{w_0}{\sqrt{2\pi}}\; - \sum_{
^{_{n=-\infty,\neq 0}}}^
{\infty} \frac{ \Pi_n}{ink_0}\; e^{in\theta},\end{equation}
with $\Pi_n$ the Fourier coefficients for $\Pi$, and $\Pi_0=0$ 
because of constraint (4.18). We have thus 
succeded in writing the metric of our model in terms of the
dynamical variables $u$, $p_u$, $v$, $p_v$, $w_0$ and $k_0$.

It is easy to check that the classical geometries described by Eqs. 
(4.26-28) are in fact invariant under a flip of sign in the momenta
$p_u$, $p_v$, and $k_0$.\footnote{This change of sign in the momenta, 
while keeping unaltered the configuration variables, can be 
interpreted as a time reversal.} We can take advantage of this 
symmetry to fix, e.g., the sign of $k_0$, without eliminating from 
our considerations any of the geometries that are allowed for our 
model. We will hence restrict $k_0$ to be positive from now on: 
$k_0\in I\!\!\!\,R^+$. 

Given this restriction, it is convenient to replace 
$(w_0,k_0)$ with a new pair of canonical variables, $(b_0,c_0)$,
whose respective domains of definition are the whole real axis,
\begin{equation} b_0=k_0w_0,\;\;\;\;\;\;\;\; c_0=\ln{k_0}.
\end{equation}
Note that
\begin{equation} \{b_0,c_0\}=1,\end{equation}
and that the point $k_0=0$, excluded from our phase space, has now 
been driven to the boundary of the domain of $c_0$ (namely, 
to $c_0=-\infty$). Notice also that, after performing the 
change (4.29), the Hamiltonian (4.21) turns out to be analytic
in all the elementary fields and variables of our reduced model.

To close this section, we will derive a formula equivalent to 
Eq. (4.28) that may be more useful in practice to compute the metric
function $w$. A straightforward calculation leads to
\begin{eqnarray} \dot{\Pi}&=&\{\Pi,\oint H_r\}=
\partial_{\theta}H_r,\\
\dot{w_0}&=&\{w_0,\oint H_r\}=-\frac{1}{k_0}
\oint H_r.\end{eqnarray}
It is worth remarking that the first of these equations implies
that the constraint (4.18) is preserved by the 
dynamical evolution of our model. From Eqs. (4.31,32), we also get 
the following relations among $w_0$ and the Fourier coefficients
of $\Pi$ and $H_r$ ($\Pi_n$ and $H_r^n$, respectively):
\begin{eqnarray} &&\frac{ \Pi_n}{in}=\int_{0}^{t} dt^{\prime} 
H_r^n(t^{\prime})+d_n,\;\;\;\;\;\;\;\;\;n\neq 0,\\
&&w_0=-\frac{\sqrt{2\pi}}{k_0}\int_0^{t}dt^{\prime}
H_r^0(t^{\prime}) +d_0,\end{eqnarray}
the $d_n$'s being constants for all integers $n$. Using these 
relations, it is not difficult to check that Eq. (4.28) can be 
rewritten as
\begin{eqnarray}w(t,\theta)\!&=&\!-\frac{\sqrt{2\pi}}{k_0}
\left[\int_0^t dt^{\prime} H_r(t^{\prime},\theta=0)
+\int_0^{\theta} d\theta^{\prime}\Pi(t,\theta^{\prime})
\right]\!\nonumber\\
&&+D,\end{eqnarray}
where $D$ is an undetermined real constant.\vspace*{.7cm}

\section{Quantization}
\setcounter{equation}{0}

We have seen that the four fields $(u,p_u,v, p_v)$ and the two 
homogeneous degrees of freedom ($b_0,c_0$) form a set of real 
elementary variables for our reduced model. They are only restricted
in that they are subject to the constraint (4.18). Since all our 
elementary fields are defined on $S^1$, we can expand them as 
Fourier series. The corresponding Fourier coefficients will be 
called $(u_n, p_u^n, v_n, p_v^n)$. These coefficients, together 
with ($b_0,c_0$), provide then an infinite set of homogeneous, 
elementary variables for our system. From the fact 
that $(p_u,p_v)$ are the momenta canonically conjugate to
$(u,v)$, we arrive at the following Poisson bracket structure:
\begin{equation} \{u_n, p_u^{m}\}=\{v_n,p_v^{m}\}=
\delta^m_{-n},\;\;\;\;\;\;\;\;\{b_0,c_0\}=1,\end{equation}
the rest of brackets being equal to zero. On the other hand, the 
reality conditions on our fields $(u,p_u,v,p_v)$ and variables 
$(b_0,c_0)$ imply that
\begin{equation} \overline{b_0\!}\,=b_0,\;\;\;\;\;\overline{c_0\!}
\,=c_0,\;\;\;\;\;\overline{g_n\!}\,=g_{-n},\;\;\;\;\;
\overline{p_g^n\!}\,=p_g^{-n},\end{equation}
where $g=u$ or $v$ and the bar denotes complex conjugation.

In order to quantize the system, we will follow the canonical 
quantization program put forward by Ashtekar [2]. We will first 
represent our elementary variables by linear operators acting on 
an auxiliary vector space. An inner product will be selected on
this space by imposing the reality conditions (5.2) as adjointness 
relations between our operators [2,17]. We will then represent the 
constraint (4.18) as an operator, $\hat{\Pi}_0$, and impose it on 
our quantum theory. From the kernel of $\hat{\Pi}_0$ and the inner 
product introduced on the auxiliary representation space, we will 
construct the Hilbert space of physical states, ${\cal H}_p$. 
Finally, we will identify the quantum observables of our reduced
model. By quantum observable, we mean any operator that
has a well-defined action on ${\cal H}_p$ and, therefore, commutes
with the only constraint of the model, $\hat{\Pi}_0$. In the rest of
this section, we will implement this quantization program, except 
for the discussion of the quantum observables, that will be 
postponed to Sec. VI. At some stages of our analysis, we will 
proceed only in a formal way; thus, our main concern is to show 
how the quantization process can be carried out, rather than obtain 
explicitly a rigorous and complete quantum theory for our model.

We begin by choosing as our auxiliary representation space the 
vector space of analytic functionals $\Psi$ of the set of variables
$\Omega\equiv(c_0, u_n, v_n)$ $[n=0,\pm 1,...]$. On this space, we 
can represent our elementary variables by the operators
\begin{eqnarray} &\hat{c}_0\Psi=c_0\;\Psi,\;\;\;\;\;\;\;\;\;
&\hat{b}_0\Psi=i \frac{\partial\Psi}{\partial c_0},\\
&\hat{g}_n\Psi=g_n\;\Psi,\;\;\;\;\;\;\;\;&\hat{p}_g^{-n}\Psi=
-i\frac{\partial\Psi}{\partial g_{n}},\end{eqnarray}
where $g$ stands again for $u$ or $v$ (and we have set $\hbar=1$).
Notice that the commutators of these operators reproduce the 
classical Poisson algebra (5.1) with the due factor of $i$:
\begin{equation} [\hat{b}_0,\hat{c}_0]=i,\;\;\;\;\;\;\;\;
[\hat{g}_n,\hat{p}_g^{-n}]=i.\end{equation}
The operators of our quantum theory are then given by
(possibly infinite) sums of products of the elementary operators 
(5.3,4) and c-number operators [2].

Using the reality conditions (5.2), we can select an inner product
on our auxiliary representation space in the following way.
We first adopt the ansatz 
\begin{equation} <\Phi,\Psi>=\int d\Omega\wedge d\overline{\Omega}
\;\rho(\Omega,\overline{\Omega})\;\overline{\Phi(\Omega)}\;
\Psi(\Omega),\end{equation}
with
\begin{eqnarray}d\Omega\wedge d\overline{\Omega}& &\equiv
\frac{i}{2}dc_0\wedge d\:\overline{\!c_0\!}\;\times\nonumber\\
& & \prod_{n} \left(\frac{i}{2} du_n
\wedge d\:\overline{\!u_n\!}\,\right)\,\prod_{m}\left(\frac{i}{2}dv_m
\wedge d\:\overline{\!v_m\!}\,\right)\end{eqnarray}
and $\rho$ a certain positive integration measure. The requirement
that the reality conditions (5.2) be realized as adjointness 
relations determines then the measure $\rho$ up to an overall 
positive constant,
\begin{equation} \rho=\delta(c_0-\overline{c_0\!}\,)
\prod_n\left[\delta(u_{-n}-\overline{u_{n}\!}\,)\,\delta(v_{-n}-
\overline{v_{n}\!}\,)\right].\end{equation}
In this formula, we have employed the notation
\begin{equation}\delta(z)\delta(\overline{z})\equiv \delta(x)
\delta(y),\;\;\;\;\;\;
\delta(z-\overline{z})\equiv\delta(y),\end{equation} 
for $z=x+iy$ any complex variable, and $x,y\in I\!\!\!\,R$.

We still have to impose the homogeneous constraint (4.18) quantum
mechanically. In order to do it, let us represent the phase space
variable $\Pi_0$ by the operator
\begin{equation} \hat{\Pi}_0=\sum_{s=1}^{\infty} s\,(\hat{X}_s^u+
\hat{X}_s^v),\end{equation}
where (for $g=u$ or $v$)
\begin{equation} \hat{X}_s^g=i(\hat{g}_s\hat{p}_g^{-s}-\hat{g}_{-s}
\hat{p}_g^{s}).\end{equation}
We note that, with the factor ordering chosen in Eq. (5.11),
$\hat{\Pi}_0$ is, at least formally, self-adjoint on the auxiliary
Hilbert space determined by the inner product (5.6-8).

The physical states of our quantum theory are those annihilated by 
the constraint operator $\hat{\Pi}_0$. In our auxiliary 
representation space, on the other hand, all quantum states $\Psi$
can be expressed as (possibly infinite) sums of functionals of the 
form
\begin{equation} P_{(k,\sigma)}=c_0^k\,\prod_n\left(u_n^{i_n}\,
v_n^{j_n}\right),\end{equation}
where $\sigma\equiv (i_n,j_n)$ [$n=0,\pm 1,...$], and $k$, $i_n$, 
and $j_n$ are non-negative integers. From Eqs. (5.10,11), it is easy
to check that the functionals $P_{(k,\sigma)}$ are (generalized) 
eigenfunctions of $\hat{\Pi}_0$, 
\begin{equation} \hat{\Pi}_0\,P_{(k,\sigma)}=N(\sigma) 
P_{(k,\sigma)}.\end{equation}
Here, $N(\sigma)$ is given by
\begin{equation} N(\sigma)=\sum_{s=1}^{\infty} s\,
(i_s+j_s-i_{-s}-j_{-s})\end{equation}
and is equal to the sum of indices of all factors $(u_n,v_n)$
(including multiplicities) appearing in $P_{(k,\sigma)}$.
For any quantum state $\Psi=\sum a_{(k,\sigma)} P_{(k,\sigma)}$,
with $a_{(k,\sigma)}$ some complex constants, we therefore get
\begin{equation}
\hat{\Pi}_0\Psi=\sum_{(k,\sigma)} a_{(k,\sigma)} N(\sigma) 
P_{(k,\sigma)}.\end{equation}
The uniqueness of the power series of the zero functional
implies then that $\Psi$ is annihilated by $\hat{\Pi}_0$ if and only
if it is a (generally infinite) linear combination of functionals
$P_{(k,\sigma)}$ whose total indices $N(\sigma)$ vanish.
The analytic functionals that satisfy this condition form a complex
vector space, whose Hilbert completion with respect to the inner
product (5.6-8) finally provides us with the Hilbert space of 
physical states, ${\cal H}_p$.

It is easy to see that this Hilbert space is actually infinite 
dimensional. For the sake of an example, let us display an infinite
set of states in ${\cal H}_p$, namely,
\begin{eqnarray} \Psi & = & P_{(k,\sigma^{\prime})}\;
\frac{1}{C}\left(\prod_{s\geq 0}
\frac{1}{2\pi A_s B_s}\right)\times\nonumber\\
& & \exp{\left\{-\frac{1}{4}
\left[\frac{c_0^2}{C^2}+\sum_{s=0}^{\infty}\left(
\frac{u_su_{-s}}{A_s^2}+\frac{v_sv_{-s}}{B_s^2}
\right)\right]\right\}} ,\end{eqnarray}
where $C$, $A_s$, and $B_s$ ($s=0,1,...$) are real constants, but
otherwise unrestricted, and $P_{(k,\sigma^{\prime})}$ is any 
polynomial of the form (5.12) (i.e, the set $\sigma^{\prime}$ 
contains only a finite number of non-vanishing elements in this
case) such that $N(\sigma^{\prime})=0$. All these states are 
analytic functionals of $\Omega$, belong to the kernel 
of $\hat{\Pi}_0$ and can be checked to be normalizable with
respect to the inner product (5.6-8).

\section{Quantum Observables and Hamiltonian}
\setcounter{equation}{0}

We turn now to the task of finding the quantum observables of our 
model. These are the operators with a well-defined action 
on the Hilbert space of physical states, ${\cal H}_p$. As they leave
${\cal H}_p$ invariant, they must (weakly) commute with the quantum
constraint of the model, $\hat{\Pi}_0$.

In our quantum theory, on the other hand, all operators are supposed
to be constructed from the elementary operators (5.3,4) by taking 
(suitable limits of) sums and products. In particular, it should 
be possible to obtain all quantum observables from 
(infinite) linear combinations of operators of the form
\begin{equation} \hat{P}_{(k,\Gamma)}= \hat{c}_0^{k_1}\, 
\hat{b}_0^{k_2}\prod_{n}\left[ \hat{u}_n^{i_n}\,\hat{v}_n^{j_n} 
(\hat{p}_u^{-n})^{l_{-n}}\,(\hat{p}_v^{-n})^{m_{-n}}\right],
\end{equation}
$k\equiv(k_1,k_2)$ and $\Gamma\equiv (i_n,j_n,l_n,m_n)$ 
[$n=0,\pm 1,...$] being two sets of non-negative integers. 

A straightforward calculation shows that
\begin{equation} [\hat{\Pi}_0,\hat{P}_{(k,\Gamma)}]= N(\Gamma)
\hat{P}_{(k,\Gamma)},\end{equation}
where
\begin{eqnarray} N(\Gamma)&=&\sum_{s=1}^{\infty}s\,
(i_s+j_s+l_{s}+m_{s}-i_{-s}\nonumber\\
&&\hspace*{1.2cm}-j_{-s}-l_{-s}-m_{-s})\end{eqnarray}
is [similarly to $N(\sigma)$ in Eq. (5.14) ] the sum of indices of 
all the elementary operators (counting multiplicities) that form 
$\hat{P}_{(k,\Gamma)}$. Therefore, $\hat{P}_{(k,\Gamma)}$ commutes 
with $\hat{\Pi}_0$ if and only if its total index $N(\Gamma)$ 
vanishes. Furthermore, from Eq. (6.2) and our comments above, 
it is possible to show that the quantum observables of our theory 
can always be expressed\footnote{Up to additive terms of the form
$\hat{X}\hat{\Pi}_0$ ($\hat{X}$ being a generic operator), which
vanish modulo the constraint $\hat{\Pi}_0$.}
as linear combinations (including the limit
of infinite sums) of operators $\hat{P}_{(k,\Gamma)}$ verifying 
$N(\Gamma)=0$.

A possible way to attain observables is the following.
We first define
\begin{equation} \hat{g}(\theta)=\sum_{n=-\infty}^{\infty}
\hat{g}_n\,\frac{e^{in\theta}}{\sqrt{2\pi}},\;\;\;\;\;
\hat{p}_g(\theta)=\sum_{n=-\infty}^{\infty}
\hat{p}_g^n\,\frac{e^{in\theta}}{\sqrt{2\pi}}.\end{equation}
Here, $g=u$ or $v$, and $\theta\in S^1$ must be regarded as a 
parameter. Suppose next that $g(c_0,b_0)$ is an analytic function
of $c_0$ and $b_0$, and $f[u(\theta),v(\theta)]$ a functional of 
only $u(\theta)$ and $v(\theta)$ (and perhaps of their derivatives
with respect to $\theta$) which is analytic in these fields. 
One can then check that, for all non-negative integers 
$l$ and $m$, the operators
\begin{equation} g(\hat{c}_0,\hat{b}_0) \oint f[\hat{u},\hat{v}]
(\hat{p}_u)^l\,(\hat{p}_v)^m\end{equation}
can be written as (infinite) sums of operators $\hat{P}_{(k,\Gamma)}$
with $N(\Gamma)=0$, so that they provide, in general, quantum
observables for our model.

So far, we have not discussed the dynamics of our quantum system. In
order to do it, we must first allow the physical states and quantum
observables to depend on a dynamical parameter $t\in I\!\!\!\,R$.
The quantum evolution is then dictated by the Schr\"odinger equation
\begin{equation} i\frac{\partial \Psi}{\partial t}(t)=
\hat{H}_r^T(t) \Psi(t),\end{equation}
where $\hat{H}_r^T$ is, by assumption, a self-adjoint observable that
represents the classical reduced Hamiltonian integrated over $S^1$, 
that is, the generator of the classical evolution, $\oint H_r$.

The self-adjointness of $\hat{H}_r^T$ implies that the quantum 
evolution is unitary, i.e., that it preserves the norm of the 
physical states of the theory. This is equivalent to say that the 
integration of the Schr\"odinger equation (6.6) leads then to a 
unitary evolution operator, $\hat{U}(t,0)$, such that 
$\Psi(t)=\hat{U}(t,0)\Psi$, $\Psi$ being the initial physical state. 
In terms of the quantum Hamiltonian, the evolution operator 
adopts the expression [21]
\begin{equation} \hat{U}(t,0)={\cal P}\left[\exp{\left(-i\int_0^t 
dt^{\prime} \hat{H}_r^T(t^{\prime})\right)}\right],\end{equation}
where ${\cal P}$ stands for the time ordering
\begin{eqnarray} &&{\cal P} [\hat{H}_r^T(t_1)\ldots\hat{H}_r^T(t_n)]
= \sum_{(\eta)} \hat{H}_r^T(t_{\eta(1)})\ldots 
\hat{H}_r^T(t_{\eta(n)}) \times \nonumber \\
& &\hspace*{1.5cm} \Theta(t_{\eta(1)}-t_{\eta(2)})\ldots 
\Theta(t_{\eta(n-1)}-t_{\eta(n)}).
\end{eqnarray}
Here, $\Theta$ is the Heaviside function and $\eta$ any permutation 
of the indices 1,$\ldots$, $n$.

From the Schr\"odinger equation and the self-adjointness of the
quantum Hamiltonian, we also arrive at the following
evolution for the matrix elements of any quantum observable 
$\hat{O}(t)$:
\begin{eqnarray} &&i\frac{d\;\,}{dt}<\Phi(t),\hat{O}(t)\Psi(t)>=
\nonumber \\
&&\hspace*{.8cm}<\Phi(t),\left([\hat{O}(t),\hat{H}_r^T(t)]+
i\partial_t\hat{O}(t)\right)\Psi(t)>,\end{eqnarray}
with $\partial_t\hat{O}(t)$ the derivative of $\hat{O}(t)$ with 
respect to its explicit dependence on the parameter $t$. We will then
say that an observable $\hat{O}(t)$ represents a constant of motion 
if it satisfies
\begin{equation} [\hat{O}(t),\hat{H}_r^T(t)]+i\partial_t\hat{O}(t)=0,
\end{equation}
so that all its matrix elements are constant in the quantum 
evolution. In this sense, it is worth pointing out that, given any 
quantum observable, $\hat{O}$,  that is explicitly\linebreak 
$t$-in\-dependent, one can generally obtain another observable that 
represents a constant of motion, namely,
\begin{equation} \hat{O}^{\prime}(t)=\hat{U}(t,0)\, \hat{O} \,
\hat{U}^{-1}(t,0),\end{equation}
where $\hat{U}^{-1}(t,0)$ is the inverse of the evolution operator.

We have thus seen that, in order to arrive at a unitary quantum
evolution and essentially complete our quantization, we are only 
left with the problem of finding a self-adjoint observable to 
represent the (integrated) classical reduced Hamiltonian of the 
model. A quantum Hamiltonian that, at least formally, satisfies 
these conditions is
\begin{equation} \hat{H}_r^T=-\hat{X}-e^{2t}\,\hat{Y},\end{equation}
\begin{eqnarray} & & \hat{X}=\sqrt{2\pi}e^{-\hat{c}_0}\oint
\left[4(\hat{p}_u)^2+\frac{1}{4}e^{-\hat{u}}
(\partial_{\theta}\hat{v})^2\right]\nonumber\\
& &=e^{-\hat{c}_0}\sum_{n=-\infty}^{\infty}\left[
4\sqrt{2\pi}\hat{p}_u^n\,\hat{p}_u^{-n}-\sum_{m=-\infty}^{\infty}
\frac{nm}{4}e^{-\hat{u}}_{-n-m}\,\hat{v}_n\hat{v}_m\right]
,\nonumber\\
& & \end{eqnarray}
\begin{eqnarray} & &\hat{Y}=\sqrt{2\pi}e^{-\hat{c}_0}
\oint\left[\frac{1}{16}(\partial_{\theta}\hat{u})^2+
e^{\hat{u}}(\hat{p}_v)^2\right]
\nonumber\\
& &=e^{-\hat{c}_0}\!\sum_{n=-\infty}^{\infty}\!\left[\! 
\frac{\sqrt{2\pi}}{16} n^2 \hat{u}_n \,\hat{u}_{-n}
+\sum_{m=-\infty}^{\infty}e^{\hat{u}}_{-n-m}\,\hat{p}_v^n\,
\hat{p}_v^m\right].\nonumber\\
&&\end{eqnarray}
In the above formulas,
\begin{equation} e^{\pm\hat{u}}_n=\oint e^{\pm\hat{u}}\,
\frac{e^{-in\theta}}{\sqrt{2\pi}}\end{equation}
and $\hat{u}$, $\hat{p}_u$, and $\hat{p}_v$ are the operators defined
in Eq. (6.4). It is clear that this quantum Hamiltonian commutes with
$\hat{\Pi}_0$, because it is a linear combination of operators of the
form (6.5). That this Hamiltonian is formally self-adjoint follows 
from the fact that it is given by a sum of products of commuting 
operators, as well as from the reality conditions 
$\hat{c}_0^{\dagger}=\hat{c}_0$, $\hat{g}_n^{\dagger}=\hat{g}_{-n}$
and $(\hat{p}_g^{n})^{\dagger}=\hat{p}_g^{-n}$ $(g=u$ or $v)$.
Finally, notice that $\hat{H}_r^T$ inherits an explicit 
dependence on the parameter $t$ from the time dependence of the 
classical Hamiltonian (4.21). 

To prove that the Hamiltonian (6.12-14) is in fact a self-adjoint
observable, it would actually suffice to show that it is densely
defined on the Hilbert space of physical states, ${\cal H}_p$.
From our discussion above, this would guarantee that $\hat{H}_r^T$
is a symmetric observable. That this Hamiltonian is self-adjoint
(or, strictly speaking, that it admits a self-adjoint extension)
would then be a consequence of the fact that there exists a
conjugation ${\cal C}$ on ${\cal H}_p$ which leaves the domain of 
$\hat{H}_r^T$ invariant and commutes with it [22]. We remind that 
a conjugation 
${\cal C}:{\cal H}_p\rightarrow {\cal H}_p$ is an anti-linear,
norm-preserving map whose square is the identity. It is not 
difficult to check that a map on ${\cal H}_p$ that satisfies the 
properties of a conjugation and commutes with our quantum 
Hamiltonian is
\begin{equation} {\cal C}\,\Psi(\Omega)=\overline{\Psi}({\cal C}
(\Omega)),\end{equation}
where $\Psi$ is any physical state and the action of ${\cal C}$
on the set of elementary variables $\Omega$ is given by
\begin{equation} {\cal C}\,c_0=c_0,\;\;\;\;\;{\cal C}\,u_n=u_{-n},
\;\;\;\;\;{\cal C}\,v_{n}=v_{-n},\end{equation}
with $n=0,\pm 1,...$

It could also happen that, instead on ${\cal H}_p$, the Hamiltonian
(6.12-14) admitted a self-adjoint extension only on a sufficiently 
large Hilbert subspace ${\cal H}_p^1\subset{\cal H}_p$. In that 
case, one could still try to restrict all considerations to that 
subspace in a consistent way, and regard ${\cal H}_p^1$ as the true
Hilbert space of physical states. Other\-wise, one would have to 
replace the operator (6.12-14) with a different quantum Hamiltonian 
that turned out to be physically acceptable in our model. Finally, 
if no such Hamiltonian could be found (and one insisted in arriving
at a unitary quantum evolution), one would have to start the
quantization over again, changing any of the choices that are
available in the construction of the quantum theory, like, e.g.,
the set of elementary operators or their representation.

\section{Conclusions and Further Comments}
\setcounter{equation}{0}

Starting with the Ashtekar formalism for Lorentzian general 
relativity in vacuum and restricting our attention to the sector of 
non-degenerate metrics, we have discussed the structure of the 
reduced phase space and the quantization of the family of Gowdy 
universes whose spatial topology is that of a three-torus.

We have first removed non-physical degrees of freedom by means of a
gauge fixing procedure. The gauge fixing conditions imposed, together
with the first-class constraints of the model, have been shown to 
form a set of second-class constraints that allow the reduction of 
the system. In this way, we have been able to eliminate all the 
constraints of the model except for one homogeneous constraint, 
$\Pi_0=0$. This constraint is the analogue of the periodicity 
condition studied by Gowdy [12], and generates the diffeomorphisms,
with spatially constant infinitesimal parameters, of the angular 
coordinate $\theta$ that does not correspond to a Killing field of 
the spacetime.

The choice of time that we have adopted is equivalent to that 
employed by Gowdy [12,16]. We have got rid of the Gauss constraints
and of the diffeomorphism constraints of the coordinates associated 
with Killing fields by requiring that some components of the 
densitized triad vanish [see Eqs. (2.8), (2.18), and (3.2)]. 
Finally, the $\theta$-coordinate diffeomorphism gauge freedom has 
been used to set the variable $K$ [given by Eq. (3.9)] equal to its 
mean value on each surface of constant time [16], i.e., to 
$K_0/\sqrt{2\pi}$. This quantity is known to be a constant of motion
of the model. We have then shown that the classical geometries with 
$K_0=0$ are not included in the family of cosmological solutions 
considered by Gowdy. Besides, provided that $K_0$ is different from 
zero, our gauge fixing conditions are consistent and well-posed.

We have found a canonical set of real elementary variables for
the phase space of our reduced model. This set is formed
by the four fields $u(\theta)$, $p_u(\theta)$, $v(\theta)$, and 
$p_v(\theta)$, and by the two homogeneous variables $w_0$ and $k_0$.
The reduced model is still subject to the homogeneous constraint
$\Pi_0=0$. On the other hand, the exclusion of the solutions with
$K_0=0$ implies that $k_0\in I\!\!\!\,R^+\cup I\!\!\!\,R^-$.
Making use of the fact that the classical geometries are invariant
under a change of sign in the momenta $p_u(\theta)$, $p_v(\theta)$,
and $k_0$, we have none the less been able to restrict all 
considerations to the case $k_0\in I\!\!\!\,R^+$ without loss of
generality. In order to attain a canonical set of elementary 
variables whose respective domains of definition are the entire 
real axis, we have then replaced $w_0$ and $k_0$ with a new 
canonical pair of variables, $(b_0,c_0)$.

In addition, we have obtained the explicit expression for the 
classical metric of the Gowdy spacetimes and determined the 
reduced Hamiltonian, $H_r$, that generates the dynamical evolution 
in our gauge-fixed model. This Hamiltonian presents an explicit 
dependence on the time coordinate, so that the reduced system 
is not conservative.

Since the fields $u(\theta)$, $p_u(\theta)$, $v(\theta)$, and
$p_v(\theta)$ are periodic functions of $\theta$, we can
expand them as Fourier series. The Fourier coefficients 
$(u_n,p_u^{-n})$ and $(v_n,p_v^{-n})$ turn out to be canonically 
conjugate pairs of homogeneous variables. Employing these Fourier
coefficients and $(b_0,c_0)$ as elementary variables, we have 
proceeded to quantize our model following the canonical program 
ellaborated by Ash\-tekar [2]. We have first represented the 
variables $b_0$, $c_0$, $u_n$, $p_u^n$, $v_n$, and $p_v^n$ 
$(n=0,\pm 1,...)$ as elementary linear operators acting on the 
vector space of analytic functionals of $c_0$, $u_n$, and $v_n$.
A unique inner product has been selected on this space by demanding
that the complex conjugation relations (5.2) (our reality conditions)
are realized quantum mechanically as adjointness relations. 
We have then represented the homogeneous constraint of our reduced
model by a linear operator, $\hat{\Pi}_0$, and determined the 
quantum states that are annihilated by it. These states, together 
with the inner product selected by the reality conditions, have 
provided us with the Hilbert space of physical states, ${\cal H}_p$.

The quantum observables of the reduced model are the operators that 
have a well-defined action on ${\cal H}_p$. In our quantum theory, 
on the other hand, a generic observable should always be given by a
suitable (possibly infinite) sum of products of elementary operators. 
Using this fact, we have been able to obtain the general form of 
the quantum observables.

We have finally introduced a dynamical evolution in our system by
imposing a Schr\"odinger equation with quantum Hamiltonian,
$\hat{H}_r^T$, representing the classical reduced Hamiltonian 
integrated over $S^1$. If one requires that the quantum evolution
be unitary, the Hamiltonian $\hat{H}_r^T$ must be a self-adjoint
observable. We have found an operator $\hat{H}_r^T$ that, 
at least formally, satisfies these conditions. Also discussed are
other still available, alternative possibilities to obtain a
Hamiltonian which would really be well-defined and self-adjoint.

In analyzing the structure of the phase space of our reduced model,
we have restricted the variable $k_0$ to be positive by taking
advantage of the symmetry of the classical geometries under a change 
of sign in the momenta $p_u(\theta)$, $p_v(\theta)$, and $k_0$, a 
transformation that can be regarded as a time reversal. Had we not
imposed this restriction, we should have split the phase space
into two disconnected parts: one for $k_0>0$, and the other
for $k_0<0$. Replacing definiton (4.29) with
\begin{equation} b_0=k_0 w_0,\;\;\;\;\;\;\;\;c_0=\ln{(-k_0)}
\end{equation}
in the sector of negative values of $k_0$ and repeating our 
quantization procedure, we would have then arrived at a quantum 
theory whose physical Hilbert space would be given by the direct 
sum of two copies of the Hilbert space of physical states 
constructed for $k_0\in I\!\!\!\,R^+$. Nevertheless, any of 
these two copies would actually provide us with an irreducible 
representation of the model as far as we do not allow 
time reversal operations.

On the other hand, although we have considered the invariance of the
Gowdy geometries under the transformations generated by the 
diffeomorphism constraints, we have in fact not discussed the 
possible symmetries under global diffeomorphisms that cannot be 
connected with the identity transformation. Diffeomorphisms 
of this kind which are compatible with the form of 
the metric (4.26) are given, e.g., by a change of orientation 
in one of the angular coordinates, or by an interchange of
the coordinates that correspond to Killing fields of the spacetime,
i.e. $\omega$ and $\nu$. In this sense, the point of view that 
we have adopted is that the coordinates $\omega$ and $\nu$
are physically distinguishable and the orientation of all spatial
coordinates fixed once and for all. 

During the completion of this paper, we have become aware of an
independent work by A. Ashtekar and M. Pierri [23], who
also study the quantization of the family of Gowdy 
universes with the spatial topology of a three-torus. In that work, 
the discussion has none the less been restricted to the case in 
which the two commuting Killing fields of the model are hypersurface
orthogonal.

\acknowledgments

The author is grateful to P. F. Gonz\'alez D\'{\i}az and N.
Manojlovi\'c for helpful conversations. He is also greatly 
thankful to A. Ashtekar for valuable comments and discussions.

\end{document}